\documentclass[12pt]{JHEP3}

\newcommand{\bea}{\begin{eqnarray}}
\newcommand{\beal}[1]{\begin{eqnarray}\label{#1}}
\newcommand{\eea}{\end{eqnarray}} 
\newcommand{\be}{\begin{equation}} 
\newcommand{\bel}[1]{\begin{equation}\label{#1}}
\newcommand{\ee}{\end{equation}}

\newcommand{\bit}{\begin{itemize}}
\newcommand{\eit}{\end{itemize}}
\newcommand{\ben}{\begin{enumerate}}
\newcommand{\een}{\end{enumerate}}

\newcommand{\req}[1]{(\ref{#1})}

\def\d{\partial}
\def\half{\frac{1}{2}}

\def\alp{\leavevmode\ifmmode {\alpha^\prime} \else ${\alpha^\prime}$ \fi}
\def\mps{M_P^2}

\title{New Solutions of the Inflationary Flow Equations}

\preprint{}

\author{Micha\l\ Spali\'nski\footnote{Email: mspal@fuw.edu.pl}\\
So\l tan Institute for Nuclear Studies\\
ul. Ho\.za 69, 
00-681 Warszawa, Polska.
}

\abstract{The inflationary flow equations are a frequently used
method of surveying the space of inflationary models. In these applications
the infinite hierarchy of differential equations is truncated in a way
which has been shown to be equivalent to restricting the set of models
considered to those characterized by polynomial inflaton potentials. This
paper explores a different method of solving the flow equations, which does
not truncate the hierarchy and in consequence 
covers a much wider class of models while retaining the practical
usability of the standard approach.}


\begin{document}

\section{Introduction}

The inflationary flow equations, introduced by 
Kinney \cite{Kinney:2002qn} following
earlier work by Hoffman and Turner \cite{Hoffman:2000ue} 
are frequently used as a means of surveying
the space of scalar 
field theories describing inflation. The
flow equations form an infinite hierarchy of ordinary differential
equations. They provide a convenient framework
for parameterizing the space of inflationary solutions of Einstein's
equations coupled to a single inflaton with canonical kinetic terms. 
These equations form the basis of Monte-Carlo reconstruction
of the inflaton potential \cite{Easther:2002rw}, as well as other studies which
explore the space of inflationary models
\cite{Verde:2005ff}-\cite{Powell:2007gu}.  

Practical applications of the flow equations involve truncating the
infinite hierarchy so as to obtain a closed set of equations which can be
solved. The procedure introduced by Kinney \cite{Kinney:2002qn} and used by all
subsequent studies\footnote{Some interesting alternative approaches to
  characterizing the space of inflationary models can be found in the paper
of Ramirez and Liddle \cite{Ramirez:2005cy}.} defines subspaces of solutions
characterized by all 
Hubble flow parameters vanishing apart from a finite number.
It was shown by 
Liddle \cite{Liddle:2003py} that this procedure corresponds to restricting
the set of 
all inflaton potentials to polynomials of order related to the number of
flow parameters allowed to assume non-zero values. 

The flow equations themselves do not make any assumptions about the
potential energy density which 
defines a specific model. Each solution of the flow equations however
corresponds to some definite potential, which can easily be obtained. Thus
subspaces of the space of all solutions correspond to definite classes of
potentials; dynamical
information enters the flow equations algorithm by the means 
chosen for truncating the hierarchy. That is the point when the class of
potentials to be scanned is determined.
 
While straightforward (and adequate for many purposes), the
truncation scheme introduced by Kinney \cite{Kinney:2002qn} and universally
employed in subsequent studies of the flow equations 
excludes some interesting models of inflation, for example those involving
exponentials of the inflaton field. Such cases arise in some supergravity or
string motivated models \cite{Kallosh:2007ig,Cline:2006hu}, and so it would
be nice to be able to broaden the 
scheme so that they could be included. One may argue that for
studying physical effects which are sensitive only to a 
limited range of inflaton values a polynomial approximation
for the potential may be all that is needed, but at least from the
theoretical point of view one would like to understand the choices
involved. Furthermore, in some cases discussed in the literature, such as models
involving potentials with sharp ``features'' 
\cite{Starobinsky:1992ts}-\cite{Kinney:2007ii}, the polynomial 
approximation is by definition unlikely to be sufficient. 

In view of the above it becomes interesting
to consider alternative schemes of dealing with the flow equations. The
purpose of this
note is to offer a method of solving the 
hierarchy, which does not set an infinite number of flow parameters to
zero. Indeed, all the flow parameters are non-zero in this approach, and
in consequence 
this method does not restrict the space of generated potentials to
polynomials in the inflaton field. The
hierarchy effectively terminates because flow parameters of higher order
are expressed 
algebraically in terms of a finite number lower order ones as a consequence
of a condition which requires that a flow parameter of some order be 
constant.  

The immediate question is then how the new set of potentials is related to
the set of polynomial potentials scanned in Kinney's procedure. It is
obvious from the new termination condition (introduced in section
\ref{newsol}) that the space of potentials accessible with the new method
is strictly a superset of the old 
one. One way to see the impact of the new scheme would be to
repeat Kinney's numerical study and see how the set of generated models
broadens the regions appearing in \cite{Kinney:2002qn} (and in following
studies of this 
type). This is clearly interesting to explore. This note however focuses on
some simple insights which can be gained by analytically solving the new
termination conditions at low levels, very much in the spirit of Liddle's
analytic solution of the original scheme \cite{Liddle:2003py}. Unlike that
case however, a 
complete analytic solution is possible only in the two lowest orders: at
higher orders one needs to resort to numerical methods.  
The analytic solutions
described in section \ref{anares} show that the new scheme, in accordance
with expectations, brings in
non-polynomial potentials. In particular, the lowest order solution 
describes power law inflation, which is outside the standard scheme since it
requires an exponential potential. The next order leads to some interesting
cases which have appeared in the literature in various contexts
\cite{deRitis:1991uv}-\cite{Kinney:2005vj}. They include   
so-called ``ultra-slow-roll inflation'' \cite{Tsamis:2003px}, which has the
second Hubble parameter $\eta_H=3$.
At higher orders one can write down some special solutions analytically,
but for practical applications numerical integration is required. One can
proceed to integrate the flow equations directly, or alternatively
integrate the 
termination condition, which is a single ordinary differential equation of
order $M+1$, where $M$ is the order of the flow parameter required to be
constant. The second approach directly gives $H(\phi)$, the Hubble
parameter as a function of the inflaton. Both options can be 
implemented in a 
straightforward manner and from a technical point of view neither requires 
anything beyond what is used 
in investigations using Kinney's truncation \cite{Kinney:2002qn}.

The inflationary flow equations, as well as their standard truncation are described in
section \ref{flow}. The new scheme is presented in section \ref{newsol} 
and some analytic solutions of the termination conditions are 
described in section \ref{anares}. Some closing comments are offered in
section \ref{conc}.


\section{Truncating the Flow Equations}
\label{flow}

The inflationary flow equations introduced in \cite{Kinney:2002qn} assume
that 
inflation is driven by a single scalar field described by an effective
action of the form
\be
S = - \int d^4x \sqrt{-g}(\half (\d\phi)^2 + V(\phi)) \ .
\ee
For spatially homogeneous field configurations 
Einstein equations reduce to
\bea
\dot{\rho} &=& - 3 H (p+\rho) \label{conserv}\\
3\mps H^2&=& \rho \label{friedman} \ ,
\eea
where
\bea
p     &=&\half \dot{\phi}^2 - V(\phi) \label{canonp}\\
\rho  &=&\half \dot{\phi}^2 + V(\phi) \label{canonrho} \ .
\eea
Here $M_P$ is the reduced Planck mass ($M_P^2=1/8\pi G$), the dot indicates
a time derivative and $H\equiv \dot{a}/a$.   

It is convenient to write these equations in first order
form, treating  
$\phi$ as the evolution parameter in place of $t$. From \req{conserv} --
\req{canonrho} it follows that  
\bel{phidot}
\dot{\phi} = -2 M_P^2 H'(\phi) \ ,
\ee
where the prime denotes a derivative with respect to $\phi$. Using this and
\req{canonrho} in \req{friedman} gives 
\bel{hj}
2 M_P^4 H'(\phi)^2 = 3\mps H^2(\phi) - V(\phi)\ .
\ee
This is the Hamilton-Jacobi form of the field
equations
\cite{Markov:1988yx}-\cite{Kinney:1997ne}.   

The fundamental indicator of inflation is the first Hubble flow 
parameter
\be
\epsilon_H = 2\mps (\frac{H'}{H})^2 \ ,
\ee
where the prime indicates a derivative with respect to the inflaton
field. The basic property of $\epsilon_H$ is that 
\be
\frac{\ddot{a}}{a} = H^2 (1-\epsilon_H) \ ,
\ee
which shows that the Universe is inflating if and only if $\epsilon_H<1$. 

The number of e-folds at some time $t$ before the end of
inflation at time $t_f$ is given by 
\bel{efolds}
N=\int_{t}^{t_f} H dt \ ,
\ee
so one has $dN=-H dt$.
This convention defines $N$ as the number of e-folds before the end of
inflation at $N=0$. Thus as time flows forward, $N$ decreases. 
From \req{efolds} it follows that 
\be
\frac{d}{dN} = -\frac{\dot{\phi}}{H} \frac{d}{d\phi} \ ,
\ee
which can be rewritten using \req{phidot} as 
\be
\frac{d}{dN} = 2\mps \frac{H'}{H} \frac{d}{d\phi} \ .
\ee
By direct computation one then finds 
\be
\frac{d\epsilon_H}{dN} = - 2\epsilon_H (\epsilon_H - \eta_H) \ ,
\ee
where
\be
\eta_H = 2\mps \frac{H''}{H}
\ee
is the second flow parameter. 

The derivative of $\eta_H$ involves the third derivative of $H$, which
motivates the introduction of another dimensionless flow
parameter. Proceeding 
in this way all higher derivatives of the Hubble
parameter appear and an infinite hierarchy of differential equations is
generated. It can be described compactly by introducing the infinite
sequence of Hubble flow parameters \cite{Kinney:2002qn} defined as
\bea
\lambda_0 &=& 2\mps (\frac{H'}{H})^2 \nonumber \\
\lambda_k &=& (2\mps)^k \frac{(H')^{k-1}}{H^k} \frac{d^{k+1}H}{d\phi^{k+1}}
\label{lamdef} \ , \qquad k\geq 1 \ ,
\eea
so that $\lambda_0=\epsilon_H$ and $\lambda_1=\eta_H$. 
The flow equations can now be written as
\bea
\frac{d\lambda_0}{dN} &=&  2\lambda_0 (\lambda_0 - \lambda_1) \\
\frac{d\lambda_k}{dN} &=&  \Big(- k\lambda_0 + (k-1)\lambda_1 \Big)
\lambda_k + 
\lambda_{k+1} \ , \qquad k\geq 1 \ . \label{floweq}
\eea
This is an infinite hierarchy of differential equations for the Hubble
flow parameters $\lambda_k$. Solutions of these equations for which 
$\lambda_0<1$ for a sufficiently long time describe inflating spacetimes of
interest in cosmology. 

It was emphasized by Liddle \cite{Liddle:2003py} that the 
flow equations 
do not reflect any
specific choice of potential, since their 
derivation does not make 
use of the Hamilton-Jacobi equation. This is consistent with their purpose,
which is to describe inflationary 
solutions for canonical scalar field theories 
without prejudice.
While the equations themselves do not involve a choice of scaler potential, 
any {\em specific} solution of \req{floweq} 
corresponds to a specific scalar field theory. This is because once
$\epsilon_H$ is found as a function of $N$ by solving the flow
equations one can calculate $H(\phi)$ (up to an overall
scale)\footnote{This also requires 
relating $N$ and $\phi$, which can be done using
$dN=2\mps\frac{H}{H'}d\phi$, which follows from \req{efolds} and
\req{phidot}.}.  
This in turn determines the scalar potential via 
the Hamilton-Jacobi equation \req{hj}:
\bel{pot}
V(\phi) = 3\mps H(\phi)^2 - 2 M_P^4 H'(\phi)^2 \ .
\ee
Thus every solution of the flow equations determines the
corresponding scalar 
potential (up to an overall energy scale).

The set of all solutions of the flow equations is identical to the set of
all solutions of the Hamilton-Jacobi equations for all choices of
$V(\phi)$. In that sense the full space of solutions does not reflect any
choice of dynamics -- it conveniently parameterizes the outcome of all the
possible choices. Choosing a class of solutions (a subspace of all
solutions) is however 
tantamount to a statement of dynamics, and this is what practical
applications of the flow equations do. 

The procedure introduced by Kinney \cite{Kinney:2002qn}, and elaborated on
by many authors, 
involves truncating the infinite hierarchy by setting, for some integer
$M$,  
\bel{trunco}
\lambda_k=0 \ ,  \qquad k\geq M \ .
\ee 
This yields a closed set of differential
equations for $\lambda_0\dots\lambda_M$. It is important to note that
truncating the flow equations is not an approximation: solutions to the
truncated set of equations are exact, but they span
a subset of all the solutions to the flow equations. Thus truncation is
equivalent to restricting the set of all possible potentials to some
subset. This fact was made explicit by Liddle \cite{Liddle:2003py} who
observed that 
the truncation condition \req{trunco} could (using \req{lamdef}) be written
as   
\bel{eqpoly}
\frac{d^{M+1}H}{d\phi^{M+1}} = 0 \ ,
\ee
which makes it plain that solutions of the flow equations are polynomials
of order $M$:
\bel{poly}
H(\phi) = \sum_{k=0}^M  a_k\phi^k \ .
\ee
Using this in \req{pot} implies that the corresponding scalar
potentials are polynomials in $\phi$ of order $2M$.  One can survey a large  
space of potentials by truncating the flow equations at a high level,
i.e. by taking $M$ large in \req{trunco}.

\section{New Solutions of the Flow Equations}
\label{newsol}

While the class of polynomial potentials 
appearing in the standard treatments of the flow equations 
may be sufficient for most practical
purposes, from a theoretical perspective it seems somewhat 
limited. 
Indeed, from
the point of view of embedding inflationary scalar field theories in string
theory it seems that this restriction is quite severe, since non-polynomial 
contributions to scalar 
potentials are quite common in that setting
\cite{Kallosh:2007ig,Cline:2006hu}. It turns out however that a 
very simple modification of the standard truncation of the flow hierarchy
significantly broadens the set of potentials covered without
introducing any significant complications relative to the standard
procedure outlined in the previous section. 

The idea is to replace the
truncation condition \req{trunco} by  
\bel{termin}
\lambda_M=\lambda 
\ee 
for some $M$, where $\lambda$ is a constant. This termination condition
closes the flow  
equations hierarchy  at level $M$  by introducing the constant $\lambda$.  
The hierarchy closes, because only the first $M$ differential equations are
non-trivial if \req{termin} is imposed. The equations at levels $M$ and
above become algebraic.  
The termination condition \req{termin} does not set the higher
order flow parameters to zero: they are instead expressed in terms of
the lower order 
parameters. For example \req{floweq} and \req{termin} imply 
\be
\lambda_{M+1} = \lambda \Big(M\lambda_0 - (M-1)\lambda_1\Big) \ .
\ee
Similar relations can be written down for higher flow parameters which
generically remain non-vanishing. 

The flow equations can be integrated as before for any choice of $M$ and
some reasonable set of 
values of $\lambda$. The original subset of inflationary model space is the
case of $\lambda=0$, so clearly all the solutions appearing in the old
approach are recovered. 

There are in fact two ways to proceed. One option is
to integrate the set of $M+1$ nontrivial flow equations. The 
alternative is to directly solve the termination
condition itself. In the case of standard truncation the solution of the
truncation condition \req{trunco} is \req{poly}. This way $H$ is obtained
directly, without going through the flow parameters, in fact circumventing
the flow equations themselves. In the case of the modified termination
condition one can proceed in the same spirit by expressing \req{termin}
using \req{lamdef} as
\bel{newode}
(2\mps)^M \frac{(H')^{M-1}}{H^M} \frac{d^{M+1}H}{d\phi^{M+1}} = 
\lambda \ . 
\ee
This is a single differential equation of order $M+1$ which replaces Liddle's
\req{eqpoly}. For $M > 1$ this equation is nonlinear, and one cannot solve
it analytically. It is however straightforward to solve numerically. For
that purpose it is convenient to write it as a system of first order
equations as follows. Introducing 
\be
H_k\equiv  \frac{d^{k+1}H}{d\phi^{k+1}} \qquad k=0,\dots, M
\ee
equation \req{newode} can be rewritten as a system of first order
differential equations: 
\beal{newsys}
\frac{d H_k}{d\phi} &=& H_{k+1} \qquad k=0,\dots,M-1 \nonumber \\
\frac{d H_M}{d\phi} &=& \frac{\lambda}{(2\mps)^M} \frac{H_0^M}{H_1^{M-1}} 
\ . 
\eea
Supplementing \req{newsys} with suitable initial conditions one can generalize
numerical computations of the type pioneered by Kinney \cite{Kinney:2002qn}
to the wider set of solutions described here. To this end one can express
initial 
values for the $H_k$ in terms of often used initial values for the standard
flow parameters.   
Indicating initial values by an over-bar one has, from the definition of
$\epsilon_H$, 
\be
\bar{H}_1 = \pm \bar{H}_0\sqrt{\frac{\bar{\epsilon}_H}{2\mps}} \ . 
\ee
The flow equations (or \req{newsys}) determine $H$ up to
an overall scale, which can be taken as $\bar{H}_0$. The choice of sign
above reflects the possibility of the inflaton rolling to the left or to
the right. 
Similarly for $k\geq 1$ one can write
\be
\bar{H}_{k+1} =  (\frac{1}{2\mps})^k \bar{H}_0\bar{\lambda}_k
\sqrt{\frac{\bar{\epsilon}_H}{2\mps}} \ ,
\ee
where $\bar{\lambda}_k$ are initial values of the flow parameters, which can
be related directly to those used by Kinney \cite{Kinney:2002qn}: one has 
$\bar{\lambda}_1 = \half (\bar{\sigma}_H + 4\bar{\epsilon}_H)$  
where $\bar{\sigma}_H$ is the initial value of Kinney's $\sigma_H$, and 
$\bar{\lambda}_k$ are the initial values of Kinney's ${ }^k\lambda_H$
(for $k\geq 1$). One can now numerically integrate the equations
\req{newsys} choosing initial values of $\epsilon_H, \sigma_H, { }^k\lambda_H$
from the same ranges as those used in \cite{Kinney:2002qn} (and most of the
literature devoted to this subject) to facilitate comparison. 

The numerical computations following from this prescription will not be
presented here; instead the following section will describe some analytic
considerations which give a glimpse of 
space of potentials defined by the procedure introduced above.

\section{Some Analytic Results}
\label{anares}

To understand the difference in the space of potentials scanned by the
truncation of the flow equations described in the last section it is
instructive to look at the simplest cases, that is when \req{termin} is
imposed with $M=0$ and $M=1$, which are very simple to solve analytically. 

In the case $M=0$ the termination condition \req{termin} is the statement
that $\lambda_0$ is constant. Since 
$\lambda_0$ is just $\epsilon_H$, this is power law inflation (when
$\lambda<1$). The 
termination condition \req{termin} becomes
\bel{plitrunc}
2\mps (\frac{H'}{H})^2 = \lambda \ .
\ee
For this to make sense one can only allow non-negative values of $\lambda$.  
As discussed in the last section, \req{plitrunc} can be
regarded (in the spirit of \cite{Spalinski:2007dv})   
as a first order differential equation for $H(\phi)$. The general
solution is
\be
H(\phi) = A \exp (\pm\sqrt{\frac{\lambda}{2\mps}} \phi) \ .
\ee
This involves one integration constant, $A$. 
The corresponding potential, obtained form \req{pot} is
\be
V(\phi) = A^2 \mps (3-\lambda) \exp(\pm\sqrt{\frac{2\lambda}{\mps}} \phi) 
\ .  
\ee
This is the well known example of Lucchin and Matarrese
\cite{Lucchin:1984yf}.  

This simplest case already shows the difference between the procedure 
proposed in the previous section and the one normally used in the
literature. While the standard procedure is equivalent to scanning over the
set of polynomial potentials of some order, here one obtains a
non-polynomial one.  
Clearly, all the Hubble flow parameters are non-zero: they are all 
given by powers of the constant $\lambda$. The 
case of standard level 0 truncation is obtained in the limit
$\lambda\rightarrow 0$,
which describes de Sitter expansion. 

Imposing the termination condition \req{termin} with $M=1$ is also
solvable, and rather interesting. The termination condition
reads
\be
2\mps \frac{H''}{H} = \lambda \ .
\ee
Here there is no restriction on the sign of $\lambda$, and the character of
the solutions of 
this differential equation depend on this sign. Since the differential
equation is of second order there will be two integration constants. 
The general solution is\footnote{Potentials of this type were previously
  considered in references \cite{deRitis:1991uv,Easther:1993qg}. They can
  all be characterized as models with constant $\eta_H$.}
\be
H(\phi) = \left\{\begin{array}{ll}
A \cosh (\sqrt{\frac{\lambda}{2\mps}}\phi) + B \sinh
(\sqrt{\frac{\lambda}{2\mps}}\phi) & \qquad \textrm{for $\lambda > 0$}\\
A + B \phi & \qquad \textrm{for $\lambda=0$}\\
A \cos (\sqrt{\frac{|\lambda|}{2\mps}}\phi) + B \sin
(\sqrt{\frac{|\lambda|}{2\mps}}\phi) & \qquad \textrm{for $\lambda < 0$ .}
\end{array}\right.
\ee
The case $\lambda=0$ is of course the result of the standard truncation at
this level. 

For a sensible inflationary solution one has to make a choice of
integration constants and restrict the range of $\phi$ appropriately so as
to ensure that $H'$ does not change sign. 
Rather than discuss this further, this presentation will focus on one
special case, that of $\lambda > 0$ with $B=0$:
\bel{cosh}
H(\phi) = A \cosh (\sqrt{\frac{\lambda}{2\mps}}\phi) \ .
\ee
The Hubble slow-roll parameters are
\bea
\epsilon_H &=& \lambda\tanh^2 (\sqrt{\frac{\lambda}{2\mps}}\phi) \\ 
\eta_H &=& \lambda \ .
\eea
The corresponding potential, which follows from \req{pot} reads
\bel{coshpot}
V(\phi) = A^2\mps  \Big( \lambda + 
(3-\lambda) \cosh^2 (\sqrt{\frac{\lambda}{2\mps}}\phi)\Big) \ . 
\ee
Note that for $\lambda=3$ the potential becomes constant: this solution was
discussed by Tsamis and Woodard \cite{Tsamis:2003px} under the name
ultra-slow-roll inflation. It was later analysed by 
by Kinney as an example where the spectrum of
curvature perturbations is exactly scale invariant but where the
horizon-crossing formalism fails \cite{Kinney:2005vj}. 
The solution \req{cosh} is valid for a range of $\lambda$, so taking  
$\lambda$ close to $3$, but not exactly $3$, should 
provide interesting examples with almost scale invariant
spectra\footnote{A very interesting and nontrivial class of potentials with
exactly scale invariant spectra was obtained recently by Starobinsky
\cite{Starobinsky:2005ab}. It 
appears that these models are not of the type considered here.}.  
Since the potential \req{coshpot} in this case is not polynomial, such
examples are outside the 
realm of standard truncated flow equation simulations. 

Terminating the hierarchy using \req{termin} with $M>1$ involves solving
a nonlinear equation of order $M+1$. The general solution, depending on
$M+1$ constants of integration appears not to be available analytically,
but one special solution is easy to write down: it is 
\be
H(\phi) = A\exp (\pm \frac{1}{\sqrt{2\mps}} \lambda^{\frac{1}{2M}} \phi) 
\ . 
\ee
Both choices of sign are admissible, but one cannot take linear
combinations, as the equation is not linear. Therefore for 
higher-level solutions one needs to resort to numerical integration, as 
discussed in the previous section.

\section{Conclusions}
\label{conc}

The inflationary flow equations are the basis of a very widely used
approach to exploring the realm of inflationary scalar field theories. The
standard method of truncating the infinite hierarchy of flow equations
restricts the class of scalar potentials to polynomials in the inflaton
field.  This paper presented a different, more general, way of solving the
hierarchy. The resulting class of potentials includes those obtained by the
standard truncation method, but is much broader in that it also includes a
wide range of non-polynomial potentials. The procedure boils down to
solving the system of differential equations \req{newsys}, which is the
main result presented here.

Some insight into the space of potentials accessible using this method can
be gleaned from analytic solutions to lowest level termination conditions.
As discussed in section \ref{anares}, examples found this way have already
appeared in the literature in various contexts. Here they serve the purpose
of illustrating how the new solutions of the flow equations extend the
range of potentials scanned.

The approach described here could be particularly useful for exploring
models of inflation involving potentials with ``features'', where the
potential is non-polynomial 
in an essential way \cite{Starobinsky:1992ts}-\cite{Kinney:2007ii}.  It
would 
also be interesting to understand the impact of the present work on
numerical flow equation simulations used to analyze data from WMAP and
other sources \cite{Peiris:2003ff,Peiris:2006ug,Kinney:2006qm}. 

As this paper was being written up the generalization of the flow equations
to the case of DBI models was introduced by Peiris et
al. \cite{Peiris:2007gz}. The new  
procedure presented here can be used without change also in that case. Due
to the occurrence of the ``Lorentz'' factor $\gamma$ in the
Dirac-Born-Infeld action, the hierarchy of flow equations 
requires two truncation conditions, which Peiris et al. 
solved, demonstrating that their method scans the space of $V(\phi)$ and
$\gamma(\phi)$ which are polynomials in the inflaton field. Replacing the
two truncation conditions used in \cite{Peiris:2007gz} by conditions of the
sort advocated here works in the same way as the canonical case discussed
in this note. 
It would clearly be of interest to investigate how modifying that study
along these lines  affects 
the results reported there, given the tight observational constraints on
that class of models.

\end{document}